\documentclass[12pt,a4paper]{article}
\usepackage[margin=2.5cm]{geometry}
\usepackage{amsmath,amssymb}
\usepackage{graphicx}
\usepackage{natbib}
\usepackage{booktabs}
\usepackage{hyperref}
\usepackage{xcolor}
\usepackage{float}

\newcommand{\um}{\ensuremath{\,\mu\text{m}}}
\newcommand{\msq}{\ensuremath{\,\text{m}^2\,\text{s}^{-1}}}
\newcommand{\Deff}{D_\text{eff}}
\newcommand{\twet}{t_\text{wet}}
\newcommand{\kdo}{k_{d0}}

\title{Martian concretion sizes predicted from two independently constrained inputs: atmospheric dust grain size and obliquity-forced wetting duration}
\author{Samuel Cody}
\date{Preprint}

\begin{document}
\maketitle

\begin{abstract}
Diagenetic concretions have now been identified at multiple widely separated sites on Mars, including locations at Meridiani Planum (Opportunity), Gale crater (Curiosity), and Jezero crater (Perseverance). A systematic compilation of published sizes (Table~\ref{tab:observed}) shows that solid concretions at all sites consistently fall within the millimetre size range (typically 1--6\,mm diameter), despite differing cement mineralogies. The one substantial outlier---centimetre-to-decimetre-scale hollow concretions on Bradbury Rise at Gale crater---formed in coarser, iron-rich basaltic sandstone rather than fine-grained dust-bearing sediment, and represents a distinct formation regime. I propose that this convergence reflects a common physical control: the globally uniform fraction of ultra-fine ($\sim$3\um), amorphous, equant atmospheric dust incorporated into sediments at all sites where solid concretions are found. I derive the diagenetic timescale from Mars' $\sim$120\,kyr obliquity cycle, which drives periodic volatile redistribution and subsurface wetting at mid-latitudes: the duration of each high-obliquity wetting pulse ($\sim$10$^4$--10$^5$\,yr) sets the available time for concretion growth. Using a diffusion--reaction model with nucleation competition and this obliquity-derived timescale, I show that the low effective diffusivity imposed by the fine dust matrix limits concretion growth to the observed millimetre scale, independent of local fluid chemistry. The model further predicts that concretion formation efficiency in dust-rich sediment exceeds 90\%, making their formation essentially inevitable wherever liquid water contacts the dust, explaining their abundance. This mechanism depends critically on the non-phyllosilicate, equant-grain mineralogy of Martian dust, which maintains connected pore networks unlike terrestrial clay-sized sediments. Concretion growth is self-limiting: the first wetting pulse exhausts the reactive phases in each concretion's depletion halo, so that successive obliquity cycles produce new concretions in fresh sediment rather than enlarging existing ones. Each concretion thus records a single wetting episode. The concretions mark a specific conjunction of conditions: post-Noachian surface aridity sufficient to generate the fines, quasi-periodic obliquity-driven subsurface wetting, and a dust mineralogy that favours diffusion-limited cementation. The narrow size distributions at all sites with dust-rich host sediment are consistent with periodic forcing of comparable duration, suggesting that Martian concretion populations may constitute a sedimentary archive of the planet's orbital history.
\end{abstract}

\section{Introduction}

One of the most startling moments in the history of Martian surface exploration occurred on 25 January 2004 (24 January local time at the Jet Propulsion Laboratory). On that date, Mars Exploration Rover~B, Opportunity, landed safely on Mars' equatorial Meridiani Planum, a site selected for its enigmatic hematite spectral signature detected from orbit by Mars Global Surveyor \citep{Christensen2000}.

When the first hazard-avoidance and navigation camera images were revealed to the world via live feed from the JPL control room, the sense of shock seemed palpable. After four landing sites in succession (Vikings~1 and~2, Pathfinder, and MER-A Spirit) that looked not too dissimilar from terrestrial deserts in volcanic regions, with a fairly even scattering of mostly basaltic boulders across an ochre plain, here at last was a place that looked utterly alien. The surface was dominated by dark, coarse-textured material of an unusual nature that was not, in those first wide-angle images, immediately identifiable. Meridiani Planum was different.

And when the first Microscopic Imager frames came down, revealing the true nature of that coarse surface texture, the shock translated for many into astonishment. The dark material responsible for the hematite signature consisted of vast numbers of small spheres---hard, grey, roughly spherical pebbles, quickly nicknamed ``blueberries'' by the science team \citep{Squyres2004}. They were all of a similar size, roughly 1--6\,mm in diameter. And they appeared to have eroded out of the underlying light-toned, fine-grained sedimentary rocks, accumulating on the surface as a lag deposit.

What were they? And why were they all the same size?

In the two decades since, concretions of similar sizes but different compositions have been found at multiple additional sites on Mars, by different rovers, in different geological settings. The mystery has deepened. In this paper, I attempt to resolve these questions---not by appeal to the specific chemistry of any one site, but through the physics of the host sediment that is common to all of them.

\subsection{Concretions across Mars}

Since the discovery of the Meridiani blueberries, diagenetic concretions have emerged as one of the most widespread and recognisable sedimentary features on Mars. Comparable structures have subsequently been identified in lacustrine mudstones at Gale crater by Curiosity \citep{Stack2014,Nachon2017,Sun2019} and in the delta deposits at Jezero crater by Perseverance \citep{Farley2022,Kalucha2024}. These occurrences span a wide range of geological settings and cement mineralogies, yet the solid concretions at all sites share a striking characteristic: they consistently fall within the millimetre diameter range.

Table~\ref{tab:observed} compiles published size data for all known Martian concretion populations. At Meridiani Planum, the hematite blueberries range from 0.1 to 6.2\,mm in diameter with a mean near 4\,mm \citep{Squyres2004,Calvin2008,Misra2014}; spherule size varies systematically with elevation in the walls of Victoria Crater, with smaller spherules at higher stratigraphic levels \citep{Calvin2008}. Compositionally distinct, iron-poor spherules up to $\sim$3\,mm diameter were later found at the Kirkwood outcrop on Endeavour Crater's rim \citep{Squyres2012}. At Gale crater, Curiosity identified millimetre-scale solid, hollow, and filled nodules in the Sheepbed mudstone at Yellowknife Bay \citep{Stack2014,Grotzinger2014}, along with diverse concretion assemblages throughout more than 300\,m of Murray formation stratigraphy \citep{Sun2019}. At Jezero crater, Perseverance documented concretions ranging from $\sim$1 to 16\,mm (median 2.65\,mm) in the Shenandoah formation \citep{Kalucha2024}, and millimetre-scale spherules at two additional localities: the Neretva Vallis inlet channel and the ``St.\ Pauls Bay'' float rock on the crater rim \citep{Jones2025}.

One population stands apart from this otherwise consistent picture. At Point Lake, Twin Cairns Island, and nearby outcrops on Bradbury Rise (Gale crater), Curiosity encountered hollow spheroids and voids ranging from $\sim$1 to 23\,cm in diameter \citep{Wiens2017}. These features have thin ($\sim$1--4\,mm) iron-rich walls and formed within dark-toned, coarse-grained basaltic sandstone with total FeO$_\text{T}$ exceeding 25\,wt\%, rather than in the fine-grained, dust-bearing sediments that host all other known Martian concretions. Their formation likely involved reaction of reduced iron with oxidising groundwater at a chemical front, producing cemented rind concretions rather than the solid, diffusion-limited structures found elsewhere \citep{Wiens2017}. These hollow concretions are discussed further in Section~\ref{sec:outlier}.

\begin{table}[H]
\centering
\caption{Published sizes of observed Martian concretions with estimated fine-dust content of the host sediment. Solid concretions in fine-grained, dust-bearing sediments cluster in the 1--6\,mm range regardless of site or cement mineralogy. The hollow concretions on Bradbury Rise, formed in coarser basaltic sandstone with lower dust content, constitute the sole outlier. The X-ray amorphous component measured by CheMin (where available) serves as a proxy for atmospheric dust incorporation, since the amorphous fraction is compositionally similar to modern aeolian dust at all Gale crater sites \citep{Dehouck2014,Achilles2020}.}
\label{tab:observed}
\footnotesize
\resizebox{\textwidth}{!}{%
\begin{tabular}{lllllr}
\toprule
Site & Rover & Size range & Host lithology & Dust-sized fines & Refs.\ \\
\midrule
\multicolumn{6}{l}{\textit{Solid concretions in fine-grained sediment}} \\[2pt]
Meridiani (Burns fm.) & Opp. & 0.1--6.2\,mm (mean $\sim$4) & Sulphate-cemented sandst. & High$^a$ & [1,2,3] \\
Meridiani, Kirkwood & Opp. & $\leq$3\,mm & Noachian outcrop & Uncertain & [4] \\
Gale, YKB (Sheepbed) & Cur. & mm-scale (solid \& hollow) & Lacustrine mudstone & Very high$^b$ & [5,6] \\
Gale, Murray fm.\ (300\,m) & Cur. & mm to low-cm & Lacustrine mudstone & Very high$^c$ & [7] \\
Jezero, Shenandoah fm.\ & Per. & 1--16\,mm (med.\ 2.65) & Deltaic siltstone/sandst. & High$^d$ & [8] \\
Jezero, Neretva Vallis & Per. & $\sim$mm-scale & Sed.\ rocks, inlet channel & Not yet characterised & [9] \\
Jezero, Witch Hazel Hill & Per. & $\sim$mm-scale & Float rock, crater rim & Not yet characterised & [9] \\[4pt]
\multicolumn{6}{l}{\textit{Hollow concretions in coarser sandstone (outlier)}} \\[2pt]
Gale, Bradbury Rise & Cur. & 1--23\,cm & Basaltic sandstone & Low$^e$ & [10] \\
\bottomrule
\multicolumn{6}{l}{\footnotesize [1]~\citet{Squyres2004}; [2]~\citet{Calvin2008}; [3]~\citet{Misra2014}; [4]~\citet{Squyres2012}; [5]~\citet{Stack2014};} \\
\multicolumn{6}{l}{\footnotesize [6]~\citet{Grotzinger2014}; [7]~\citet{Sun2019}; [8]~\citet{Kalucha2024}; [9]~\citet{Jones2025}; [10]~\citet{Wiens2017}.} \\[3pt]
\multicolumn{6}{l}{\footnotesize $^a$Fine aeolian sandstone with pervasive dust admixture; sediment composition overlaps global dust \citep{Berger2016}.} \\
\multicolumn{6}{l}{\footnotesize $^b$Grains below MAHLI resolution ($< 14\um$); X-ray amorphous component 27--31\,wt\% \citep{Dehouck2014}.} \\
\multicolumn{6}{l}{\footnotesize $^c$Predominantly mudstone with grains below camera resolution; amorphous 15--73\,wt\% (mean $\sim$40\,wt\%) \citep{Achilles2020}.} \\
\multicolumn{6}{l}{\footnotesize $^d$Concretions restricted to finer-grained (siltstone--fine sandstone) units; Fe/Mg-sulphate cement 20--30\,wt\% \citep{Kalucha2024}.} \\
\multicolumn{6}{l}{\footnotesize $^e$Dark-toned sandstone with unresolved but coarser grains; FeO$_\text{T} > 25$\,wt\%; no CheMin drill data \citep{Wiens2017}.}
\end{tabular}}
\end{table}

A consistent pattern emerges from this compilation: every locality where solid, mm-scale concretions occur is hosted in sediment with a substantial component of dust-sized fines (Table~\ref{tab:observed}). At Gale crater, the CheMin X-ray diffractometer provides a quantitative proxy: the X-ray amorphous fraction, compositionally similar to modern Martian atmospheric dust \citep{Dehouck2014}, averages $\sim$40\,wt\% in Murray formation mudstones and 27--31\,wt\% in the Sheepbed mudstone \citep{Achilles2020}. These values likely underestimate the total fine-dust contribution because some originally amorphous dust may have recrystallised during diagenesis. At Meridiani, the Burns formation is an aeolian sandstone whose bulk geochemistry closely tracks global dust composition in S, Cl, and Fe enrichment \citep{Berger2016}, indicating pervasive dust incorporation during deposition. At Jezero, the concretion-bearing Shenandoah units are specifically the finest-grained facies (siltstone to fine sandstone), with coarser intervals lacking concretions \citep{Kalucha2024}. Conversely, the Bradbury Rise basaltic sandstone that hosts the anomalously large hollow concretions is texturally coarser, lacks CheMin drill data to quantify its amorphous content, and formed through a different diagenetic mechanism (Section~\ref{sec:outlier}). This pattern, mm-scale solid concretions exclusively in dust-rich sediment, larger and morphologically distinct features where dust is subordinate, is precisely what the model predicts.

Previous studies have largely treated each site's concretions individually, focusing on local geochemistry and depositional environment. Here I take a different approach, asking: why are the solid concretion sizes so similar across sites?  I propose that the answer lies not in the chemistry of the cements but in the physics of the host sediment; specifically, the globally uniform fraction of ultra-fine atmospheric dust that is incorporated into sedimentary deposits across Mars. The Bradbury Rise hollow concretions, which formed in a lithologically distinct and coarser host, provide a natural control case that reinforces this interpretation.

Martian atmospheric dust has been compositionally and granulometrically homogenised by billions of years of global circulation \citep{Lemmon2015}. It is predominantly in the low single-digit micrometre range ($\sim$3\um{} mean diameter), composed of amorphous basaltic glass, nanophase iron oxides, and poorly crystalline silicates \citep{Morris2006,Berger2016}. Critically, unlike terrestrial clay-sized particles, which are dominantly platy phyllosilicates with swelling fabric, Martian dust grains are roughly equant and non-swelling. This distinction has profound implications for pore network connectivity and fluid transport.

\section{Model description}

\subsection{Diagenetic timescale from obliquity forcing}

I begin by establishing the timescale over which subsurface liquid water is available for diagenesis, as this is the critical parameter controlling concretion size. Rather than treating this as a free variable, I derive it from the orbital mechanics of Mars.

Mars experiences three principal orbital variations: eccentricity ($\sim$95\,kyr and $\sim$2.4\,Myr periods), precession of the longitude of perihelion ($\sim$51\,kyr), and obliquity ($\sim$120\,kyr). Of these, obliquity---the axial tilt---exerts the dominant control on volatile redistribution \citep{Ward1973,Laskar2004}. Unlike Earth, where the Moon stabilises axial tilt within a narrow range (22.1$^\circ$--24.5$^\circ$), Mars lacks a large stabilising satellite. Its obliquity swings between approximately 15$^\circ$ and 45$^\circ$ on the $\sim$120\,kyr cycle, and chaotic secular resonances may have driven excursions well beyond 60$^\circ$ in the geological past \citep{Touma1993,Laskar2004}.

These oscillations drive a volatile redistribution cycle. At high obliquity, polar insolation increases dramatically, destabilising polar ice deposits. Sublimated water vapour is redistributed to lower latitudes as ground ice or surface frost. General circulation models predict that at obliquities above $\sim$35$^\circ$, substantial ice deposits accumulate at mid-latitudes \citep{Mischna2003,Forget2006}. As obliquity subsequently decreases, these deposits become thermodynamically unstable. At depth, where overburden pressure and geothermal heat are sufficient, the ice melts---producing transient liquid water or concentrated brines that infiltrate the surrounding sediment.

The duration of each wetting pulse is set by the time the obliquity spends above the critical threshold for ice mobilisation ($\sim$35$^\circ$). For the regular $\sim$120\,kyr oscillation, this is approximately 10$^4$--10$^5$ years. Prolonged high-obliquity epochs during chaotic excursions may extend wetting to $\sim$10$^6$ years. I therefore adopt $\twet \approx 10^5$ years as the reference diagenetic timescale, with an envelope of 10$^4$--10$^6$ years reflecting the range of possible obliquity forcing durations. This is a prediction of orbital mechanics, not a fitted parameter.

\subsection{Diffusion-limited concretion growth}

Given the obliquity-derived wetting timescale, I model concretion growth as a diffusion-limited process. A growing concretion draws dissolved species inward from a surrounding depletion halo. In the spherically symmetric case, the concretion radius evolves as:
\begin{equation}\label{eq:Rdiff}
R_\text{diff} = \sqrt{2\,\Deff\,\Omega\,t}
\end{equation}
where $\Deff$ is the effective diffusivity in the porous matrix, $\Omega$ is a dimensionless supersaturation parameter, and $t$ is the obliquity-derived wetting timescale (Section~2.1). The effective diffusivity is given by:
\begin{equation}\label{eq:Deff}
\Deff = \frac{D_0\,\phi}{\tau^2\,R_f}
\end{equation}
where $D_0$ is the free-solution diffusivity ($\sim$10$^{-9}$\msq), $\phi$ is the porosity (0.35), $\tau$ is the tortuosity (1.6), and $R_f$ is a retardation factor that accounts for reversible adsorption of dissolved species onto grain surfaces. Crucially, $R_f$ scales inversely with grain diameter $d$:
\begin{equation}\label{eq:Rf}
R_f = 1 + \frac{\rho_s\,(1-\phi)}{\phi}\cdot\frac{\kdo}{d}
\end{equation}
where $\rho_s$ is the grain density (2800\,kg\,m$^{-3}$) and $\kdo$ is a surface-area-normalised sorption parameter (10$^{-6}$\,m). For Martian dust at $d \approx 3\um$, $R_f$ reaches $\sim$500, suppressing diffusivity by nearly three orders of magnitude relative to a coarse sandstone. This strong retardation is the primary mechanism limiting concretion size in dust-rich sediments.

\subsection{Nucleation competition}

A second constraint arises from the competition between neighbouring concretions for dissolved species. In a reactive, fine-grained matrix, nucleation sites are abundant. I parameterise the nucleation density as $n(d) = n_0/d$, giving a characteristic inter-nucleation spacing $L_\text{nuc} = (d/n_0)^{1/3}$. Each concretion can grow to at most half this spacing before its depletion halo impinges on that of its neighbours:
\begin{equation}\label{eq:Rnuc}
R_\text{nuc} = \frac{1}{2}\left(\frac{d}{n_0}\right)^{1/3}
\end{equation}

The predicted concretion radius is the minimum of the diffusion limit and the nucleation spacing limit: $R_\text{max} = \min(R_\text{diff},\,R_\text{nuc})$. The model identifies a crossover grain size ($\approx$17\um) below which diffusion is the controlling limit and above which nucleation spacing dominates.

The scaling $n(d) = n_0/d$ can be derived from grain-contact arguments. For a random packing of spheres of diameter~$d$, the number density of grains is $N_g = 6(1-\phi)/(\pi d^3)$. With a coordination number $Z \approx 6$ for random packing, the total number of grain--grain contacts per unit volume is $N_\text{contacts} = Z\,N_g/2 \propto 1/d^3$. If the fraction of contacts that serve as active nucleation sites scales with the contact area ($\propto d^2$), then the active nucleation density is $n \propto (1/d^3)\cdot d^2 = 1/d$, recovering the assumed scaling.

Alternatively, this scaling follows from a surface-area argument: the total grain surface area per unit volume is $S_v = 6(1-\phi)/d$, and if nucleation probability is proportional to available surface, we again obtain $n \propto 1/d$.

\subsection{Formation efficiency}

I define a capture efficiency $\eta$ representing the fraction of mobilised solute that precipitates in concretions rather than being lost to advective flushing. Using the Kozeny--Carman permeability ($k \propto d^2\phi^3/180(1-\phi)^2$) and a local P\'eclet number $\text{Pe} = v \cdot R_\text{max}/\Deff$, I write:
\begin{equation}\label{eq:eta}
\eta = \frac{1}{1+\text{Pe}}
\end{equation}

This gives $\eta \to 1$ when diffusion dominates (fine grains, low permeability) and $\eta \to 0$ when advective flushing is rapid (coarse grains, high permeability). This functional form is a zeroth-order Damk\"ohler number estimate that captures the correct asymptotic scaling in both limits \citep{Steefel1994,Lichtner1996}. For Mars dust at $d = 3\um$ and the reference pressure gradient of 1\,Pa\,m$^{-1}$, Pe~$\approx 10^{-3}$--$10^{-1}$, placing the system deep in the diffusion-dominated regime where the exact functional form of the transition is irrelevant and $\eta > 0.9$.

The Darcy velocity $v$ depends on the hydraulic gradient $dP/dL = \rho_w g\,(dh/dL)$, where $dh/dL$ is the dimensionless head gradient and $g = 3.72$\,m\,s$^{-2}$ on Mars. The obliquity wetting mechanism (Section~2.1) supplies water through ice-melt infiltration into flat-lying sedimentary deposits, not through artesian springs or channelised flow, so the relevant gradients are set by local topographic slopes in the depositional settings where concretions are observed. For the flat plains at Meridiani ($dh/dL \approx 10^{-5}$), $dP/dL \approx 0.04$\,Pa\,m$^{-1}$; for typical low-gradient sedimentary terrain such as the Gale crater floor ($dh/dL \approx 10^{-3}$), $dP/dL \approx 4$\,Pa\,m$^{-1}$; and for moderately sloped terrain near crater walls or delta fronts at Jezero ($dh/dL \approx 3\times10^{-2}$), $dP/dL \approx 110$\,Pa\,m$^{-1}$. I therefore adopt a reference gradient of 1\,Pa\,m$^{-1}$ with an envelope of 0.1--100\,Pa\,m$^{-1}$, spanning the full range of plausible depositional settings. At the reference gradient, $\eta > 0.9$ for Mars dust at 3\um.

\subsection{Parameter sensitivity and independent constraints}\label{sec:sensitivity}

The retardation factor $R_f$ (Eq.~\ref{eq:Rf}) is the dominant control on effective diffusivity, and the surface-area-normalised sorption parameter $\kdo = 10^{-6}$\,m therefore requires independent justification.

\subsubsection{Sorption parameter validation}

For equant spheres of diameter~$d$ and density~$\rho_s$, the geometric specific surface area is $S_a = 6/(\rho_s\,d)$. At $d = 3\um$ and $\rho_s = 2800$\,kg\,m$^{-3}$, this gives $S_a \approx 714$\,m$^2$\,kg$^{-1}$. The volumetric partition coefficient is related to $\kdo$ through $K_d = \kdo \cdot S_a$, giving:
\begin{equation}
K_d = 10^{-6}\,\text{m} \times 714\,\text{m}^2\,\text{kg}^{-1} = 7.1 \times 10^{-4}\,\text{m}^3\,\text{kg}^{-1} \approx 0.71\,\text{L\,kg}^{-1}
\end{equation}

This value can be compared with independently measured partition coefficients for the relevant dissolved species on relevant substrates. For Fe(II) and Fe(III) sorption onto hydrous ferric oxide (ferrihydrite), the surface complexation database of \citet{Dzombak1990} implies $K_d$ values in the range $10^{-4}$--$10^{-1}$\,L\,kg$^{-1}$ depending on pH and solution composition. Batch sorption experiments on Columbia River basalt and its amorphous alteration products give $K_d \approx 10^{-4}$--$10^{-2}$\,L\,kg$^{-1}$ for Fe and Sr \citep{Ames1980}. For Ca(II) sorption on amorphous aluminosilicates, \citet{Davis1990} report comparable ranges. The derived $K_d \approx 0.7$\,L\,kg$^{-1}$ sits within the independently measured range, at the higher end consistent with the high specific surface area and reactive amorphous mineralogy of Mars dust.

\subsubsection{Sensitivity to $\kdo$ and $n_0$}

\begin{figure}[H]
\centering
\includegraphics[width=\textwidth]{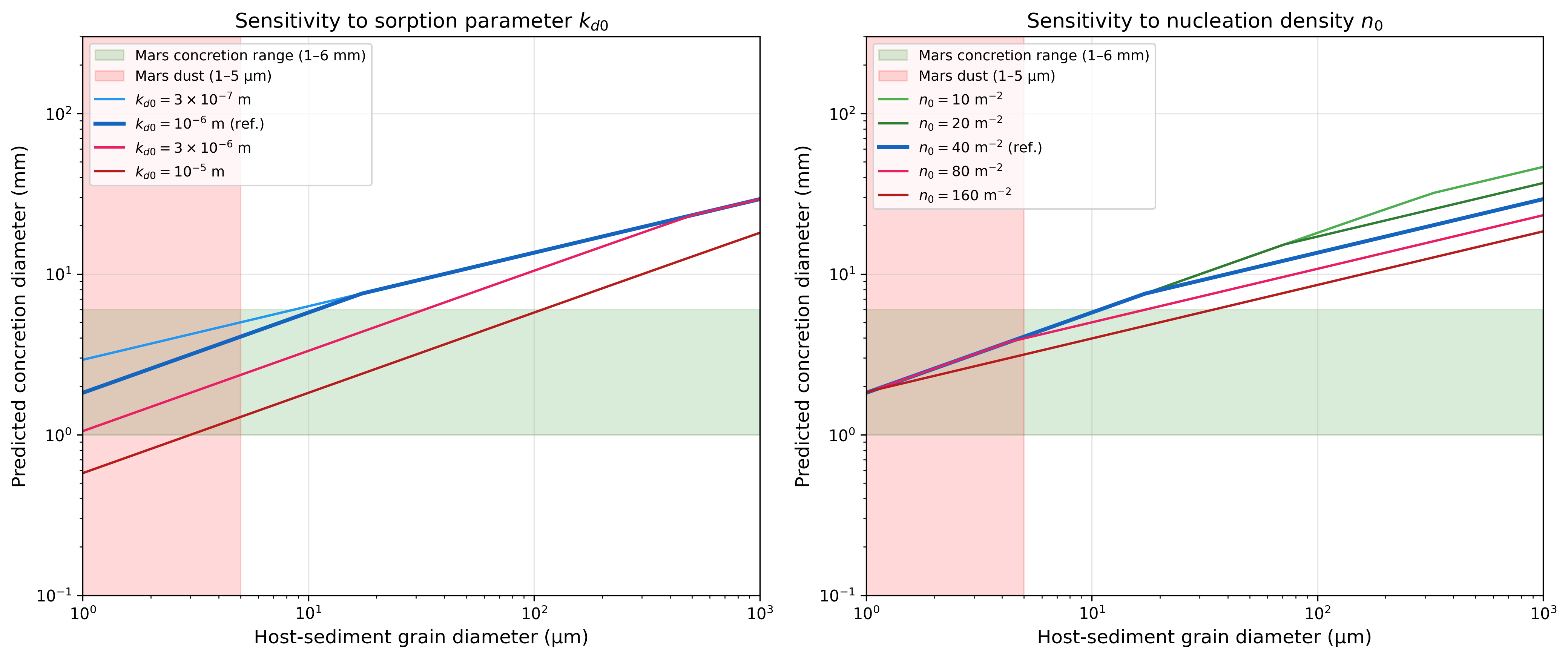}
\caption{Sensitivity of predicted concretion diameter to (\textit{left}) the sorption parameter $\kdo$ across two orders of magnitude and (\textit{right}) the nucleation density coefficient~$n_0$. Green band: observed range for solid Martian concretions in fine-grained sediment (1--6\,mm diameter; Table~\ref{tab:observed}). At 3\um{} grain size (red dashed line), the prediction remains within the observed range across the full plausible parameter space. The hollow Bradbury Rise concretions (1--23\,cm) plot far above this band, consistent with formation in coarser host sediment (Section~\ref{sec:outlier}).}
\label{fig:sensitivity}
\end{figure}

Figure~\ref{fig:sensitivity} shows the predicted concretion diameter as a function of host-sediment grain size for $\kdo$ spanning two orders of magnitude ($3\times10^{-7}$ to $10^{-5}$\,m). At the Mars dust grain size of 3\um, the predicted diameter ranges from $\sim$0.6\,mm to $\sim$6\,mm across this full range, spanning the observed envelope at all Martian sites with dust-rich host sediment (Table~\ref{tab:observed}) without requiring a specific tuned value. The prediction is robust because $R_\text{diff} \propto 1/\sqrt{R_f}$ and $R_f \propto \kdo/d$, so concretion size scales as $\sqrt{d/\kdo}$: an order-of-magnitude change in $\kdo$ changes the predicted size by only a factor of $\sim$3.

The sensitivity to the nucleation density coefficient $n_0$ is even weaker. At $d = 3\um$, the diffusion limit ($R_\text{diff} \approx 1.6$\,mm) is reached well before the nucleation spacing limit for any physically reasonable value of~$n_0$ (Figure~\ref{fig:sensitivity}, right panel). This means the model prediction at the Mars dust grain size is controlled entirely by diffusion and is insensitive to the nucleation model. The nucleation density only becomes relevant for grain sizes above the crossover ($\sim$17\um{} at the reference parameters), where it sets the maximum size in coarser sediments.

\subsubsection{Nucleation density: observational constraint on the scaling exponent}

\begin{figure}[H]
\centering
\includegraphics[width=\textwidth]{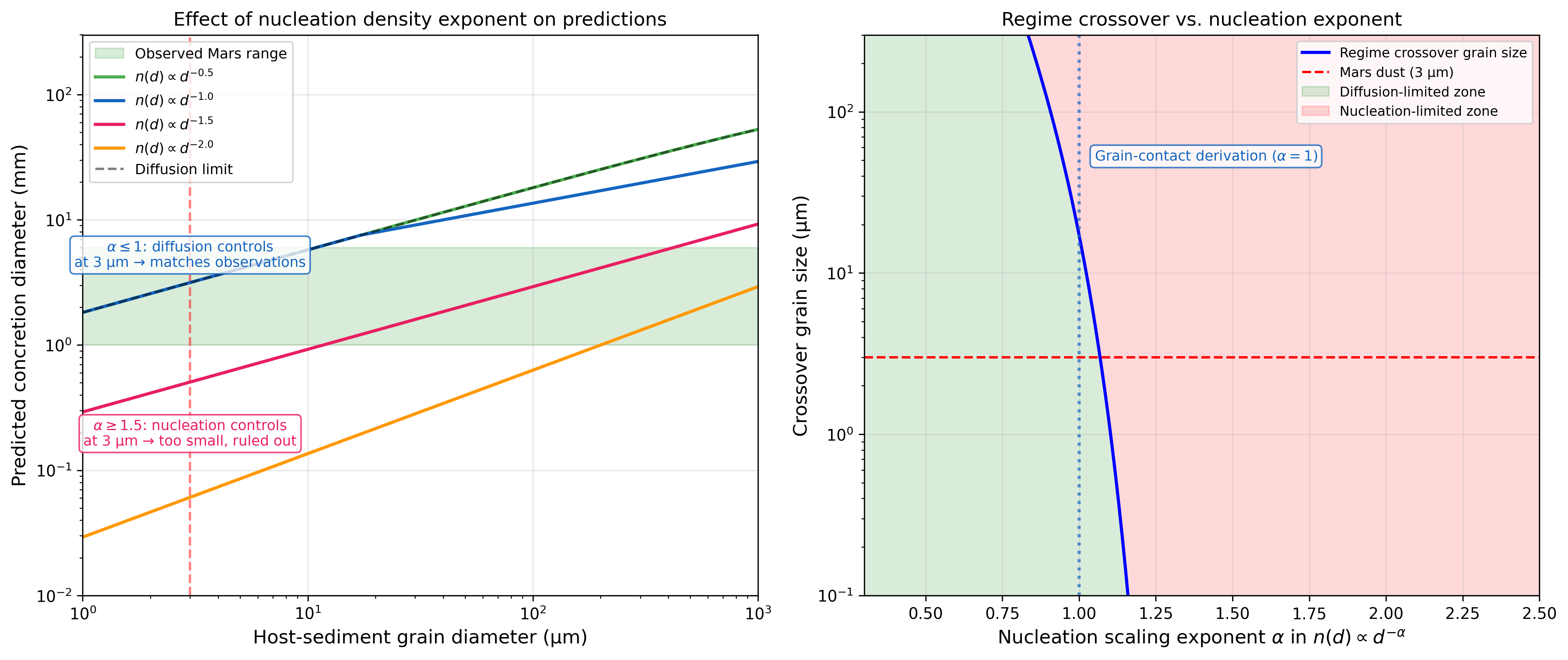}
\caption{(\textit{Left}) Predicted concretion diameter for different nucleation density scaling exponents~$\alpha$. For $\alpha \leq 1$, diffusion controls at the Mars dust grain size and predictions match observations (green band). For $\alpha \geq 1.5$, nucleation dominates even at 3\um, predicting concretions well below the observed range, ruled out by the data. (\textit{Right}) Crossover grain size as a function of~$\alpha$ (log scale). Green region: diffusion-limited zone (crossover above Mars dust at 3\um). Pink region: nucleation-limited zone. The grain-contact derivation (dotted line, $\alpha = 1$) sits within the diffusion-limited zone, consistent with observations.}
\label{fig:nucleation}
\end{figure}

The insensitivity to the nucleation scaling can be demonstrated more generally. Figure~\ref{fig:nucleation} shows that for nucleation density scalings $n(d) \propto d^{-\alpha}$ with $\alpha$ ranging from 0.5 to 2.0, all predictions converge at $d = 3\um$ because the system is deep in the diffusion-limited regime. The crossover grain size between the diffusion and nucleation regimes shifts with~$\alpha$, but for any $\alpha \lesssim 1.1$, Mars dust at 3\um{} remains diffusion-limited.

The grain-contact derivation (Section~2.3) gives $\alpha = 1$, and this value is arguably an upper bound. Alternative assumptions about the loading of nucleation sites onto grain surfaces yield smaller exponents: if nucleation requires not merely surface area but geometrically favourable concavities (such as grain-contact dimples), the fraction of surface that is active decreases with decreasing grain size, giving $\alpha < 1$. Similarly, if nucleation is limited by the availability of chemically distinct impurity phases rather than by geometric contacts, $n(d)$ may scale more weakly with~$d$ or become approximately constant, corresponding to $\alpha \to 0$. In all such cases, the prediction at the Mars dust grain size remains diffusion-limited and unchanged.

This insensitivity breaks down at higher exponents. For $\alpha \geq 1.5$, nucleation spacing becomes the controlling limit even at 3\um, and the predicted concretion diameter drops to $\sim$0.5\,mm or smaller, well below the observed 1--6\,mm range at all Martian sites (Table~\ref{tab:observed}). The observed sizes thus independently constrain the nucleation exponent: $\alpha \lesssim 1.1$ is required for consistency with the data. The grain-contact derivation ($\alpha = 1$) satisfies this constraint; alternative loading assumptions give even smaller values; and steeper scalings ($\alpha \geq 1.5$) are ruled out. The conclusion is therefore robust against substantial uncertainty in the nucleation scaling. This provides a rare example of a microphysical parameter being observationally constrained by macroscopic concretion populations on another planet.

\section{Results}

\subsection{Predicted concretion sizes}

Table~\ref{tab:sizes} summarises the model predictions across the Wentworth grain-size scale. At Mars dust grain sizes ($\sim$3\um), the model predicts concretion diameters of $\sim$3\,mm in the diffusion-limited regime, consistent with the 1--6\,mm range observed for solid concretions in fine-grained sediment at all Martian sites (Table~\ref{tab:observed}). For comparison, medium sandstone ($\sim$250\um) predicts concretions of $\sim$18\,mm, and coarse sand ($\sim$1\,mm) predicts $\sim$29\,mm, broadly consistent with the larger concretions found in terrestrial sandstones such as the Navajo Sandstone of Utah \citep{Chan2004,Chan2005}.

\begin{table}[H]
\centering
\caption{Predicted concretion diameter as a function of host-sediment grain size.}
\label{tab:sizes}
\begin{tabular}{rlrl}
\toprule
Grain $\varnothing$ (\um) & Sediment class & Concretion $\varnothing$ (mm) & Regime \\
\midrule
1 & Clay & 1.8 & Diffusion \\
3 & Mars dust (typ.) & 3.2 & Diffusion \\
5 & Very fine silt & 4.1 & Diffusion \\
10 & Fine silt & 5.8 & Diffusion \\
20 & Medium silt & 7.9 & Nucleation \\
63 & Coarse silt & 11.6 & Nucleation \\
125 & Fine sand & 14.6 & Nucleation \\
250 & Medium sand & 18.4 & Nucleation \\
500 & Coarse sand & 23.2 & Nucleation \\
1000 & Very coarse sand & 29.2 & Nucleation \\
\bottomrule
\end{tabular}
\end{table}

Figure~\ref{fig:threepanel}A shows the predicted concretion diameter as a function of host-sediment grain size, with the hollow Bradbury Rise concretions and terrestrial analogues plotted for comparison. The model reproduces the observed sizes at all Martian sites with dust-rich host sediment within the obliquity-derived wetting timescale envelope of 10$^4$--10$^6$ years. The hollow Bradbury Rise concretions (open circle), plotted at the inferred fine-to-very-fine sand range ($\sim$60--200\um; grains unresolved by MAHLI), fall above the mm-scale cluster but within the model's predicted range for fine sandstone, consistent with the coarse-host interpretation developed in Section~\ref{sec:outlier}. Figure~\ref{fig:threepanel}B compares the observed concretion sizes at all seven dust-rich Martian localities directly against the model prediction at $d = 3\um$: all fall within the predicted envelope.

\subsection{Formation efficiency and the clay paradox}

\begin{figure}[H]
\centering
\includegraphics[width=\textwidth]{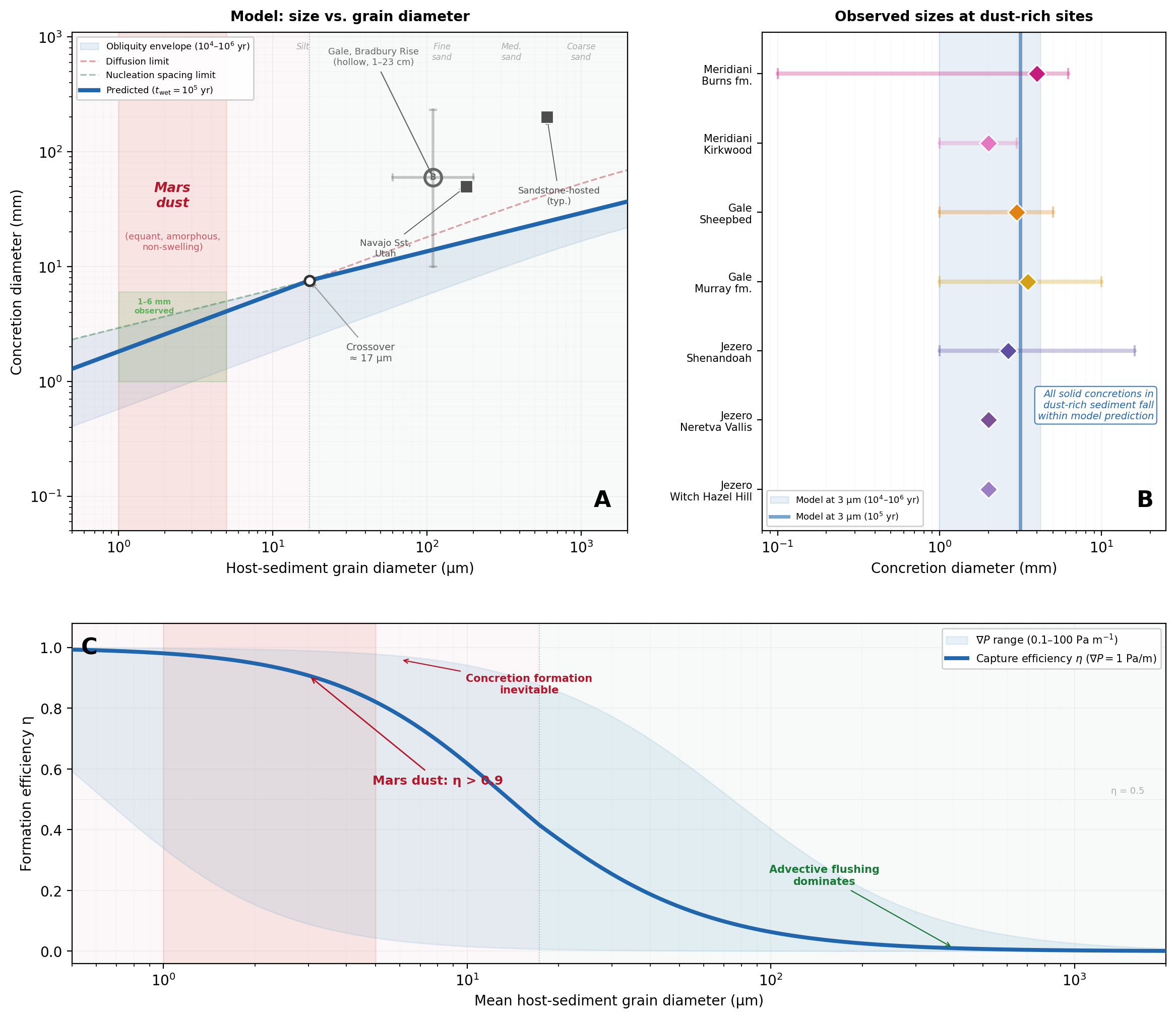}
\caption{(\textbf{A})~Predicted concretion diameter vs.\ host-sediment grain size. Blue curve: model prediction for the reference wetting timescale $\twet = 10^5$\,yr. Red dashed: diffusion limit. Green dashed: nucleation spacing limit. Shaded blue envelope: range of obliquity forcing durations ($10^4$--$10^6$\,yr). Open circle with horizontal error bar: hollow Bradbury Rise concretions plotted at the inferred fine-to-very-fine sand range ($\sim$60--200\um); filled squares: terrestrial analogues. The Mars dust band (red shading) overlaps the clay-size range but, unlike terrestrial phyllosilicate clays, maintains connected pore networks that permit diffusion-limited cementation. Green box: observed 1--6\,mm size range of solid Martian concretions. (\textbf{B})~Observed concretion sizes at all seven dust-rich Martian localities (diamonds with range bars where published) compared with the model prediction at $d = 3\um$ (vertical blue line and shaded envelope). All solid concretions fall within the predicted range. (\textbf{C})~Concretion formation efficiency~$\eta$ as a function of host grain size. Blue curve: capture efficiency at the reference pressure gradient ($\nabla P = 1$\,Pa\,m$^{-1}$). Shaded envelope: range of site-specific topographic gradients (0.1--100\,Pa\,m$^{-1}$). At the Mars dust grain size, $\eta > 0.9$: concretion formation is essentially inevitable.}
\label{fig:threepanel}
\end{figure}

Figure~\ref{fig:threepanel}C pairs the size prediction with formation efficiency. At Mars dust grain sizes, the capture efficiency exceeds 90\%, meaning that concretion formation is essentially inevitable wherever liquid water contacts dust-rich sediment. This efficiency drops steeply through the silt range, falling below 2\% at medium sand sizes, where advective flushing dominates and concretion formation becomes contingent on specific hydrological conditions.

The figure also addresses the apparent paradox that terrestrial clay-sized sediments rarely contain concretions. I show that the model's predictions in the clay-size range apply only to equant, non-swelling grains such as Martian dust. Terrestrial phyllosilicate clays have platy grain morphologies that produce swelling fabric with bound interlayer water, effectively disconnecting the pore network and preventing the diffusive transport that drives concretion growth. Mars dust is clay-sized but mineralogically distinct: amorphous basaltic glass and nanophase iron oxides with equant grain shapes that maintain connected pore networks. The Kozeny--Carman permeability model is applicable to Mars dust but not to terrestrial phyllosilicate clays. Mars dust thus occupies a grain-size/mineralogy combination with essentially no terrestrial analogue---and it is precisely this combination that makes it an exceptionally efficient concretion factory.

The uniformity of dust grain size and equant grain morphology also explains the characteristically spherical shape of Martian concretions. In an isotropic porous medium, the depletion halo surrounding a growing concretion expands symmetrically, producing spherical growth. Martian atmospheric dust, being both narrow in its size distribution and equant in grain shape, approximates an isotropic diffusion medium. Terrestrial clay-sized sediments, by contrast, have platy grain morphologies with strong fabric alignment that would produce anisotropic diffusivity and correspondingly irregular or elongated concretion shapes even if the pore network remained connected. The near-spherical morphology observed at Meridiani, Gale, and Jezero is thus a geometric consequence of the same dust properties that control concretion size.

\section{Discussion}

\subsection{A physical rather than chemical control}

The central finding of this work is that the millimetre size range of Martian concretions is set by the physical transport properties of the host sediment rather than by the specific chemistry of the precipitating cement. The globally uniform Martian dust fraction imposes a common effective diffusivity across sites, and this diffusivity limits concretion growth to a narrow and predictable size range. The cement mineralogy (hematite at Meridiani, calcium sulphate at Gale, iron-bearing phases at Jezero) fills in whichever geochemical system happens to be locally available, but it does not control the dimensions of the resulting concretions.

This decoupling of size from chemistry explains an observation that has not been adequately addressed in the literature: why missions to multiple geologically distinct sites, spanning sulphate-cemented aeolian sandstone, lacustrine mudstone, and deltaic siltstone, all found solid concretions of essentially the same size (Table~\ref{tab:observed}).

The parameter sensitivity analysis (Section~\ref{sec:sensitivity}) reinforces this conclusion. The key parameters divide into two categories: those that are well-constrained independently ($d$, $D_0$, $\phi$, $\tau$, $\rho_s$, $\twet$) and those with greater uncertainty ($\kdo$, $n_0$, $\Omega$). The model's prediction at Mars dust grain sizes is robust to the uncertain parameters because the system sits deep in the diffusion-limited, low-Pe regime where these parameters enter only weakly (Section~\ref{sec:sensitivity}). The prediction is controlled primarily by grain size and its effect on retardation, both of which are measured quantities.

\subsection{Concretion abundance as a consequence of formation efficiency}

The formation efficiency analysis (Figure~\ref{fig:threepanel}C) provides a quantitative explanation for the ubiquity of Martian concretions. In the diffusion-limited regime occupied by dust-rich sediments, over 90\% of dissolved species are captured into concretions. The process is self-organising: low permeability traps pore fluids, high reactive surface area ensures rapid dissolution and local supersaturation, and steep chemical gradients create a strong thermodynamic drive pulling ions towards nucleation sites. It is almost thermodynamically difficult \emph{not} to form concretions under these conditions.

In coarser terrestrial sediments, by contrast, formation efficiency drops below a few per cent. Here, the matrix is permeable enough for advective flow to flush pore fluids through before local supersaturation develops. Concretion formation becomes contingent on the right combination of flow rates, geochemical conditions, and timing. The diffusion-limited regime produces concretions almost inevitably; the nucleation-limited regime produces them only when conditions happen to be right.

The $\eta = 1/(1+\text{Pe})$ formulation is a standard Damk\"ohler-type scaling that correctly captures the balance between diffusive capture and advective loss \citep{Steefel1994,Lichtner1996}. For Martian dust at Pe~$\ll 1$, the exact functional form is immaterial: any reasonable interpolation between the diffusion-dominated and advection-dominated limits gives $\eta > 0.9$. A quasi-static spherical diffusion model, included as supplementary material, confirms that the diffusive flux into a growing concretion in a 3\um-grain matrix overwhelmingly exceeds advective loss at the pressure gradients relevant to flat-lying Martian sediments.

\subsection{The Bradbury Rise hollow concretions: a coarse-host outlier}\label{sec:outlier}

The centimetre-to-decimetre-scale hollow concretions at Point Lake, Twin Cairns Island, and nearby outcrops on Bradbury Rise (Gale crater) are the one clear departure from the otherwise tight mm-scale size clustering seen across Mars (Table~\ref{tab:observed}). Rather than contradicting the dust-diffusion model, however, these features reinforce it by illustrating the consequences of a different host lithology.

The Bradbury Rise concretions differ from the solid concretions at all other sites in three key respects. First, their host rock is a dark-toned, iron-rich basaltic sandstone with FeO$_\text{T} > 25$\,wt\% \citep{Wiens2017}, mapped as the ``rugged unit'' on Bradbury Rise \citep{Jacob2014}. Individual grains in this material were not resolved by MAHLI, but the rock's texture, composition, and stratigraphic position within the Bradbury Group, which consists of fluvial conglomerates, cross-bedded sandstones, and mudstones \citep{Grotzinger2014}, indicate a fine to very fine sandstone, substantially coarser than the $\sim$3\um{} atmospheric dust that dominates the concretion-bearing sediments at Meridiani, the Sheepbed mudstone, the Murray formation, and the Jezero delta deposits.

Second, these concretions are hollow, with walls only 1--4\,mm thick \citep{Wiens2017}. This morphology is unlike the solid, cement-filled concretions at all other Martian sites and points to a fundamentally different growth mechanism: precipitation of iron-oxide cement as a rind at a chemical reaction front, rather than progressive infilling of pore space by diffusion-limited cementation.

Third, the favoured formation model involves oxidation--reduction chemistry rather than simple dissolution and reprecipitation. \citet{Wiens2017} proposed that reduced iron in the host sandstone reacted with oxidising groundwater, producing an outward-migrating reaction front at which Fe$^{2+}$ was oxidised and reprecipitated as an iron-oxide rind. This mechanism, analogous to terrestrial ``liesegang''-type or iron-curtain concretions formed around pyrite or siderite nodules in porous sandstones, can produce structures at the centimetre-to-decimetre scale because the controlling length scale is set by the distance over which the redox front propagates, not by the diffusion-limited depletion halo of a nucleation site.

In the framework of the present model, the Bradbury Rise host sandstone would sit well above the crossover grain size ($\sim$17\um), in the regime where retardation is low ($R_f \ll 100$), effective diffusivity is high, and neither diffusion limitation nor nucleation competition constrains growth to the millimetre scale. At fine-to-very-fine sand grain sizes ($\sim$60--200\um), Table~\ref{tab:sizes} predicts concretion diameters of 12--18\,mm, already an order of magnitude larger than in dust-rich sediment. The hollow-rind morphology further relaxes the size constraint because cement need not fill the entire volume, only a thin shell. The coarser host also implies lower formation efficiency ($\eta \ll 0.9$), consistent with the localised occurrence of these features at only a few outcrops rather than the pervasive distributions seen in dust-bearing mudstones.

The Bradbury Rise concretions thus serve as a natural control case: where the host sediment is coarser and the formation mechanism differs, the resulting concretions are correspondingly larger and morphologically distinct. The tight mm-scale clustering observed everywhere else on Mars is specific to the fine-grained, dust-rich sediments in which the diffusion--retardation mechanism described in this paper operates.

\subsection{Hydrological implications: a chronometric constraint}

The concretions encode a specific and sequential set of conditions in Martian history. First, a prolonged period of surface aridity was required to generate the fine dust through aeolian erosion and atmospheric comminution, and to distribute it globally. This is not early Noachian Mars with its thicker atmosphere and possible standing water; this is a Mars that has already dried out substantially at the surface.

Second, liquid water had to be reintroduced into the subsurface, but without the energy to erode or rework the dust into coarser fluvial sediments. The mechanism was most likely groundwater rise, ice-melt infiltration, brine migration, or deliquescence: processes that saturate the pore space of already-deposited, dust-rich sediment from below or within.

Third, the system had to remain wet long enough for diffusion-limited growth to reach the millimetre scale (the model suggests 10$^4$--10$^6$ years), but not so energetically that advective flushing destroyed the high capture efficiency.

This temporal sequence, dust generation on an arid surface followed by quiet, persistent subsurface wetting, is most consistent with late Hesperian to early Amazonian conditions. Intriguingly, this suggests that the concretion-forming episodes at Meridiani, Gale, and Jezero may be broadly coeval, even though the sites are separated by thousands of kilometres.

The concretions are thus arguably a more specific hydrological indicator than many commonly cited features. Channels and deltas record surface water flow; phyllosilicates record water--rock interaction. But concretions in dust-rich sediment specifically record the presence of liquid water in the subsurface of an already-arid Mars---a low-energy, diffusion-dominated regime of quiet, persistent moisture rather than catastrophic flooding. And their mm-scale size quantitatively constrains the transport properties of that environment.

\subsection{Concretions as an obliquity proxy}

Because the diagenetic timescale in the model is derived from obliquity forcing (Section~2.1) rather than fitted to observations, the predicted concretion sizes constitute an independent test of the obliquity--volatile coupling hypothesis. The agreement between the predicted $\sim$3\,mm diameter at $\twet \approx 10^5$ years and the observed sizes at all sites with dust-rich host sediment (Table~\ref{tab:observed}) is consistent with concretion growth being paced by the $\sim$120\,kyr obliquity cycle.

This connection has a deeper implication. Because Mars' obliquity evolution is chaotic on timescales beyond $\sim$20\,Myr \citep{Laskar2004}, backward integration of the orbital elements cannot reliably reconstruct the obliquity history of early Mars. However, if concretion size is set by the duration of obliquity-driven wetting pulses, then the size distribution of concretions at a given site encodes information about the statistics of those pulses. The narrow size distributions observed at Meridiani, Gale, and Jezero, across seven distinct concretion-bearing localities (Table~\ref{tab:observed}), would indicate that the wetting events were of comparable duration, consistent with quasi-periodic obliquity forcing rather than singular or stochastic events. Concretion size distributions thus offer a potential sedimentary proxy for the obliquity history of Mars; a record that is otherwise inaccessible through dynamical reconstruction.

The tight clustering of concretion sizes also constrains the forcing regime. If the available wetting time were highly variable (as it would be under stochastic or chaotically dominated forcing), one would expect a correspondingly wide spread in concretion diameters, since $R_\text{diff} \propto \sqrt{t}$. The narrow spread suggests a repeatable, quasi-periodic driver, exactly what the regular $\sim$120\,kyr obliquity oscillation provides. Conversely, sites preserving concretions with anomalously large or bimodal size distributions could indicate prolonged or irregular obliquity excursions, potentially marking transitions between stable and chaotic obliquity regimes.

One observed trend is particularly suggestive of a depth-dependent wetting control. At Victoria Crater, Opportunity found that spherule diameter decreases systematically with increasing elevation (i.e.\ shallower stratigraphic position) in the crater walls, with the largest blueberries near the bottom and the smallest near the top \citep{Calvin2008}. A similar pattern has been reported in the Murray formation at Gale crater, where \citet{Sun2019} documented an upsection decrease in overall concretion size through more than 300\,m of stratigraphy. \citet{Sun2019} attributed this trend to decreasing porosity and permeability upsection. The obliquity-forcing model offers a complementary and perhaps more fundamental explanation.

Sediments at greater burial depth are thermally buffered from the surface: diurnal, seasonal, and even multi-annual temperature oscillations are damped exponentially with depth, with an $e$-folding skin depth of order $\sim$1\,m for seasonal cycles and $\sim$10\,m for obliquity-period cycles in a low-thermal-inertia regolith. During an obliquity-driven wetting event, ice melting at depth thus experiences a more stable thermal environment. Deeper sediments remain above the melting threshold for a larger fraction of the high-obliquity interval, giving a longer effective wetting time $\twet$ and consequently larger concretions ($R_\text{diff} \propto \sqrt{\twet}$). Shallower sediments, by contrast, are exposed to larger and more rapid temperature excursions. Wetting in near-surface layers would be more intermittent; freeze--thaw cycling would repeatedly interrupt diffusion-limited growth, and the cumulative effective wetting duration would be shorter. The result is smaller concretions at shallower stratigraphic levels, exactly as observed.

This mechanism predicts a continuous, monotonic relationship between burial depth at the time of diagenesis and concretion diameter, with the steepest gradient in the upper few metres where thermal buffering changes most rapidly. The relationship should be independent of cement chemistry but sensitive to the thermal diffusivity of the overburden. If the size--depth trend can be measured quantitatively at Victoria Crater or in the Murray formation, it would provide a direct constraint on the thermal skin depth of Hesperian-age Martian sediment, and by extension on the thermal inertia and volatile content of the near-surface at the time of concretion formation.

Finally, the obliquity connection may explain the apparent coevality of concretion formation across geographically distant sites. If all dust-bearing sites experienced the same obliquity-driven wetting pulse (or the same phase of a quasi-periodic cycle), then concretion formation would have been triggered simultaneously across the planet. The concretions would not be recording independent local events but a single global forcing signal expressed through local sediment and fluid chemistry.

\subsection{Self-limiting growth: why concretions do not enlarge across successive cycles}

An obvious objection to the obliquity-pacing model is that if dozens of obliquity cycles have occurred, concretions should have grown incrementally across multiple wet pulses, reaching sizes far larger than those observed. I argue that concretion growth in dust-rich sediment is fundamentally self-limiting, so that each concretion records a single wetting episode rather than cumulative diagenetic history.

During the initial wet pulse, each growing concretion scavenges dissolved species from a surrounding depletion halo---a shell of matrix perhaps 5--10\,mm in radius from which the soluble amorphous dust phases have been dissolved and reprecipitated as stable crystalline cement within the concretion. When the next obliquity-driven wet pulse arrives, this halo is chemically exhausted: the reactive phases are gone, the pore space is partly occluded by cement from the first cycle, and the concretion's own surface is now a stable crystalline phase (hematite, sulphate, or equivalent) rather than reactive amorphous glass. There is neither a source of dissolved species nor a thermodynamic drive for further precipitation.

Resupply of the depletion halo from more distant, undepleted matrix is impeded by the same low diffusivity that limited the initial growth. The characteristic diffusion length for ion transport during a single wet pulse is of order $\sqrt{\Deff \cdot \twet} \approx$ a few millimetres, comparable to the concretion radius itself. To replenish the halo, ions would need to diffuse inward from beyond the depleted zone, a distance that grows with each cycle while the available time per cycle remains fixed. The geometry is self-defeating: the resupply front retreats faster than diffusion can follow.

This self-limiting behaviour is specific to the diffusion-dominated regime of fine-grained, reactive sediments. In a coarse sandstone with high permeability, advective flow can flush fresh solute through the matrix, replenishing the supply zone and permitting incremental growth across multiple fluid events. But in dust-rich sediment, the combination of low permeability and high initial reactivity ensures that the local supply is consumed in a single pulse and cannot be replenished.

Consequently, each obliquity cycle produces concretions in fresh, undepleted sediment, typically a new layer of dust deposited during the intervening dry phase, rather than enlarging existing ones. This explains both the narrow size distribution (every concretion experienced exactly one effective growth episode) and the prediction of discrete concretion-bearing horizons separated by barren intervals, each horizon recording a single wet pulse.

\section{Testable predictions}

The model generates several predictions testable by current and future missions:

\begin{enumerate}
\item Dust-rich Martian sediments that lack concretions are predicted to have experienced a different formation history, one without the low-energy subsurface wetting episodes that the obliquity-forced model requires. The absence of concretions in dust-bearing deposits would thus indicate either that the sediment was emplaced during a period of sustained low obliquity (when mid-latitude ground ice was not being mobilised), that burial was insufficient to permit ice-melt infiltration, or that the sediment was deposited or reworked under high-energy conditions that prevented the quiet, diffusion-dominated pore fluid regime necessary for concretion growth. Because Martian atmospheric dust is globally ubiquitous, the absence of concretions in dust-rich sediment is a hydrological signal, not a sedimentological one.

\item Concretions in especially dust-rich deposits should not be substantially smaller than $\sim$1--2\,mm, because nucleation spacing sets a floor on concretion size even as diffusivity continues to decrease.

\item The size distribution of concretions at any given site should be narrow (low coefficient of variation), reflecting the uniform transport properties of the dust and the single-pulse growth limitation.

\item Sites with evidence of high-energy fluid flow (e.g., strong advective flushing) should have fewer or absent concretions, even if the sediment is dust-rich, because high P\'eclet numbers suppress capture efficiency.

\item If concretion growth is paced by obliquity-driven wetting, then sites preserving multiple stratigraphic horizons of concretions should show quasi-periodic spacing between concretion-bearing layers, corresponding to the $\sim$120\,kyr obliquity period. Additionally, the variance in concretion size within a single horizon should be small relative to the variance between horizons formed under different obliquity regimes.

\item Within a single concretion-bearing stratigraphic section, concretion diameter should increase monotonically with burial depth at the time of diagenesis, reflecting the longer effective wetting duration in thermally buffered deeper sediment. The size--depth gradient should be steepest in the upper few metres and insensitive to cement mineralogy. Quantitative measurement of this trend, already qualitatively observed at Victoria Crater \citep{Calvin2008} and in the Murray formation \citep{Sun2019}, would constrain the thermal skin depth and near-surface thermal inertia of Hesperian-age Mars.

\item Future identification of concretions in sediments of independently characterised grain size provides a direct test of the model's central prediction (Table~\ref{tab:sizes}). In particular, concretions in coarse-grained sandstones, like the hollow Bradbury Rise features, should be systematically larger and morphologically distinct (hollow rinds rather than solid infill) compared with those in fine-grained, dust-rich deposits. This size--lithology correlation is predicted by the model but has not yet been systematically tested across Mars.

\item Concretion sphericity should correlate positively with the dust fraction of the host sediment. In formations where the dust fraction is borderline or unevenly distributed, the effective diffusivity becomes spatially heterogeneous: some concretions grow in locally dust-rich pockets with near-isotropic transport, while others encounter grain-size variations that introduce directional asymmetry. This heterogeneity should manifest simultaneously in both shape and size---irregular concretions should co-occur with broader size distributions within the same horizon, because the same patchiness in dust content that breaks spherical symmetry also breaks the uniformity of the diffusion-limited growth constraint. Conversely, where the dust fraction is highest and most uniform, both size and shape distributions should be narrow. The tightest clustering in both observables is predicted for the most dust-rich mudstones, while formations near the threshold dust fraction should show the greatest scatter in both. This correlation between size variance and shape variance within a single horizon would be diagnostic of the dust-diffusion control and difficult to explain by chemistry alone. The hollow Bradbury Rise concretions, which formed in dust-poor sandstone and exhibit both irregular morphologies and a wide size range (1--23\,cm), are qualitatively consistent with this prediction.

\item In stratigraphic sections with widely varying depositional rates, concretions should preferentially occur in thin, fine-grained horizons corresponding to periods of low clastic sediment flux, where the continuous background of atmospheric dustfall constitutes a higher fraction of the total deposit. Intervals dominated by fluvial, deltaic, or aeolian sand transport would dilute the dust component below the threshold required for diffusion-limited cementation, and should be barren of solid concretions. The concretion-bearing horizons would thus pick out the quietest depositional intervals within a mixed sequence, independent of cement mineralogy. The Murray formation at Gale crater, which preserves more than 300\,m of lithologically varied lacustrine stratigraphy with documented concretion occurrences at multiple levels \citep{Sun2019}, may already contain the data needed to test this prediction.
\end{enumerate}

\section{Conclusions}

I have shown that a diffusion--reaction model, combining the known physical properties of Martian atmospheric dust with a diagenetic timescale derived from Mars' $\sim$120\,kyr obliquity cycle, quantitatively predicts the mm-scale size range of solid concretions observed in dust-rich sediment at Meridiani Planum, Gale crater, and Jezero crater without fitted parameters (Table~\ref{tab:observed}). The sole substantial outlier---hollow concretions up to 23\,cm at Bradbury Rise---formed in coarser basaltic sandstone through a distinct redox-front mechanism, reinforcing the model's central prediction that concretion size is controlled by host-sediment grain size. The model identifies two independent physical controls: the globally uniform, ultra-fine, equant Martian dust sets the effective diffusivity, while the obliquity-driven volatile cycle sets the available wetting time. Together, these constrain concretion size to a narrow range independent of local cement chemistry. The prediction is robust across two orders of magnitude in the sorption parameter $\kdo$ (Section~\ref{sec:sensitivity}), and the observed concretion sizes independently constrain the nucleation density exponent to $\alpha \approx 1$, consistent with the grain-contact derivation.

The high formation efficiency ($\eta > 0.9$) in dust-rich sediments explains why concretions are ubiquitous on Mars: their formation is essentially inevitable wherever liquid water contacts dust-bearing sediment. The required hydrological sequence, surface aridity producing the fines followed by obliquity-paced subsurface wetting, places the concretions in a specific window of Martian history, most likely the late Hesperian to early Amazonian. The narrow size distributions observed at all sites with dust-rich host lithologies (Table~\ref{tab:observed}) are consistent with quasi-periodic forcing of comparable duration, suggesting that Martian concretion populations constitute a sedimentary archive of the planet's obliquity history; a record that is otherwise inaccessible due to the chaotic nature of Mars' long-term orbital evolution. The Bradbury Rise hollow concretions, formed in coarser host rock through a distinct mechanism (Section~\ref{sec:outlier}), demonstrate that where the dust-diffusion control is absent, neither the mm-scale size constraint nor the high formation efficiency applies.

\bibliographystyle{apalike}

\begin{thebibliography}{99}

\bibitem[Achilles et~al.(2020)]{Achilles2020}
Achilles, C.N., et~al., 2020. Evidence for multiple diagenetic episodes in ancient fluvial-lacustrine sedimentary rocks in Gale crater, Mars. Journal of Geophysical Research: Planets, 125, e2019JE006295.

\bibitem[Ames \& McGarrah(1980)]{Ames1980}
Ames, L.L. and McGarrah, J.E., 1980. Basalt radionuclide distribution coefficient determinations, FY1979 annual report. RHO-BWI-C-67, Rockwell Hanford Operations.

\bibitem[Berger et~al.(2016)]{Berger2016}
Berger, J.A., et~al., 2016. A global Mars dust composition refined by the Alpha-Particle X-ray Spectrometer in Gale crater. Geophysical Research Letters, 43, 67--75.

\bibitem[Calvin et~al.(2008)]{Calvin2008}
Calvin, W.M., et~al., 2008. Hematite spherules at Meridiani: Results from MI, Mini-TES, and Pancam. Journal of Geophysical Research, 113, E12S37.

\bibitem[Chan et~al.(2004)]{Chan2004}
Chan, M.A., Beitler, B., Parry, W.T., Orm\"o, J. and Komatsu, G., 2004. A possible terrestrial analogue for haematite concretions on Mars. Nature, 429, 731--734.

\bibitem[Chan et~al.(2005)]{Chan2005}
Chan, M.A., Beitler, B.B., Parry, W.T., Orm\"o, J. and Komatsu, G., 2005. Red rock and red planet diagenesis: comparison of Earth and Mars concretions. GSA Today, 15, 4--10.

\bibitem[Christensen et~al.(2000)]{Christensen2000}
Christensen, P.R., et~al., 2000. Detection of crystalline hematite mineralization on Mars by the Thermal Emission Spectrometer: Evidence for near-surface water. Journal of Geophysical Research, 105, 9623--9642.

\bibitem[Davis \& Kent(1990)]{Davis1990}
Davis, J.A. and Kent, D.B., 1990. Surface complexation modeling in aqueous geochemistry. Reviews in Mineralogy and Geochemistry, 23, 177--260.

\bibitem[Dehouck et~al.(2014)]{Dehouck2014}
Dehouck, E., et~al., 2014. Constraints on abundance, composition, and nature of X-ray amorphous components of soils and rocks at Gale crater, Mars. Journal of Geophysical Research: Planets, 119, 2640--2657.

\bibitem[Dzombak \& Morel(1990)]{Dzombak1990}
Dzombak, D.A. and Morel, F.M.M., 1990. Surface Complexation Modeling: Hydrous Ferric Oxide. Wiley, New York.

\bibitem[Farley et~al.(2022)]{Farley2022}
Farley, K.A., et~al., 2022. Aqueously altered igneous rocks sampled on the floor of Jezero crater, Mars. Science, 377, eabo2196.

\bibitem[Forget et~al.(2006)]{Forget2006}
Forget, F., et~al., 2006. Formation of glaciers on Mars by atmospheric precipitation at high obliquity. Science, 311, 368--371.

\bibitem[Grotzinger et~al.(2014)]{Grotzinger2014}
Grotzinger, J.P., et~al., 2014. A habitable fluvio-lacustrine environment at Yellowknife Bay, Gale crater, Mars. Science, 343, 1242777.

\bibitem[Jacob et~al.(2014)]{Jacob2014}
Jacob, S.R., et~al., 2014. Stratigraphy and mineralogy of Bradbury Rise, Gale crater. 45th Lunar and Planetary Science Conference, abstract 1395.

\bibitem[Jones(2025)]{Jones2025}
Jones, A., 2025. Shocking spherules! NASA Mars Exploration Program Blog, 24 March 2025. Available at: \url{https://science.nasa.gov/blog/shocking-spherules/}.

\bibitem[Kalucha et~al.(2024)]{Kalucha2024}
Kalucha, H., et~al., 2024. Probable concretions observed in the Shenandoah formation of Jezero crater, Mars and comparison with terrestrial analogs. Journal of Geophysical Research: Planets, 129, e2023JE008138.

\bibitem[Laskar et~al.(2004)]{Laskar2004}
Laskar, J., et~al., 2004. Long term evolution and chaotic diffusion of the insolation quantities of Mars. Icarus, 170, 343--364.

\bibitem[Lemmon et~al.(2015)]{Lemmon2015}
Lemmon, M.T., et~al., 2015. Dust aerosol, clouds, and the atmospheric optical depth record over 5 Mars years of the Mars Exploration Rover mission. Icarus, 251, 96--111.

\bibitem[Lichtner(1996)]{Lichtner1996}
Lichtner, P.C., 1996. Continuum formulation of multicomponent-multiphase reactive transport. Reviews in Mineralogy, 34, 1--81.

\bibitem[Mischna et~al.(2003)]{Mischna2003}
Mischna, M.A., et~al., 2003. On the orbital forcing of Martian water and CO$_2$ cycles: A general circulation model study with simplified volatile schemes. Journal of Geophysical Research, 108, 5062.

\bibitem[Misra et~al.(2014)]{Misra2014}
Misra, A.K., et~al., 2014. Possible mechanism for explaining the origin and size distribution of Martian hematite spherules. Planetary and Space Science, 92, 1--12.

\bibitem[Morris et~al.(2006)]{Morris2006}
Morris, R.V., et~al., 2006. M\"ossbauer mineralogy of rock, soil, and dust at Meridiani Planum, Mars. Journal of Geophysical Research, 111, E12S15.

\bibitem[Nachon et~al.(2017)]{Nachon2017}
Nachon, M., et~al., 2017. Chemistry of diagenetic features analysed by ChemCam at Pahrump Hills, Gale crater, Mars. Icarus, 281, 121--136.

\bibitem[Squyres et~al.(2004)]{Squyres2004}
Squyres, S.W., et~al., 2004. In situ evidence for an ancient aqueous environment at Meridiani Planum, Mars. Science, 306, 1709--1714.

\bibitem[Squyres et~al.(2012)]{Squyres2012}
Squyres, S.W., et~al., 2012. Ancient impact and aqueous processes at Endeavour Crater, Mars. Science, 336, 570--576.

\bibitem[Stack et~al.(2014)]{Stack2014}
Stack, K.M., et~al., 2014. Diagenetic origin of nodules in the Sheepbed member, Yellowknife Bay formation, Gale crater, Mars. Journal of Geophysical Research: Planets, 119, 1637--1664.

\bibitem[Steefel \& Lasaga(1994)]{Steefel1994}
Steefel, C.I. and Lasaga, A.C., 1994. A coupled model for transport of multiple chemical species and kinetic precipitation/dissolution reactions with application to reactive flow in single phase hydrothermal systems. American Journal of Science, 294, 529--592.

\bibitem[Sun et~al.(2019)]{Sun2019}
Sun, V.Z., et~al., 2019. Late-stage diagenetic concretions in the Murray formation, Gale crater, Mars. Icarus, 321, 866--890.

\bibitem[Touma \& Wisdom(1993)]{Touma1993}
Touma, J. and Wisdom, J., 1993. The chaotic obliquity of Mars. Science, 259, 1294--1297.

\bibitem[Ward(1973)]{Ward1973}
Ward, W.R., 1973. Large-scale variations in the obliquity of Mars. Science, 181, 260--262.

\bibitem[Wiens et~al.(2017)]{Wiens2017}
Wiens, R.C., et~al., 2017. Centimeter to decimeter hollow concretions and voids in Gale crater sediments, Mars. Icarus, 289, 144--156.

\end{thebibliography}

\end{document}